\documentclass[usenatbib]{mn2e}
\usepackage{graphicx}
\title{The unusual morphology of the intragroup medium in NGC\,5171}

\author[J. P. F. Osmond, T. J. Ponman \& A. Finoguenov]
  {John P. F. Osmond,$^1$\thanks{E-mail: jpfo@star.sr.bham.ac.uk}
   Trevor J. Ponman$^1$ and Alexis Finoguenov$^2$\\
  $^1$School of Physics and Astronomy, The University of Birmingham,
  Edgbaston, Birmingham, B15 2TT, UK\\
  $^2$Max-Planck-Institut f{\"u}r extraterrestrische Physik,
  Giessenbachstrasse, 85748, Garching, Germany\\}

\date{Accepted 2004 ??. Received 2004 ??; in original form 2004 ??}

\pagerange{\pageref{firstpage}--\pageref{lastpage}}

\pubyear{2004}

\newcommand{\ks}{\ensuremath{\mbox{ks}}}
\newcommand{\XMM}{\emph{XMM-Newton}}
\newcommand{\LX}{\ensuremath{L_{\mathrm{X}}}}
\newcommand{\ergps}{\ensuremath{\mbox{erg~s}^{-1}}}
\newcommand{\TX}{\ensuremath{T_{\mathrm{X}}}}
\newcommand{\kev}{\ensuremath{\mbox{keV}}}
\newcommand{\Zsol}{\ensuremath{\mathrm{Z_{\odot}}}}
\newcommand{\Z}{\ensuremath{Z}}
\newcommand{\ROSAT}{\emph{ROSAT}}
\newcommand{\sigmav}{\ensuremath{\sigma_{\mathrm{v}}}}
\newcommand{\kmps}{\ensuremath{\mbox{km~s}^{-1}}}
\newcommand{\PSPC}{\emph{PSPC}}
\newcommand{\GEMS}{\emph{GEMS}}
\newcommand{\betaspec}{\ensuremath{\beta_{\mathrm{spec}}}}
\newcommand{\kmpspMpc}{\ensuremath{\mbox{km~s}^{-1}~\mbox{Mpc}^{-1}}}
\newcommand{\Hzero}{\ensuremath{H_{\mathrm{0}}}}
\newcommand{\Mpc}{\ensuremath{\mbox{Mpc}}}
\newcommand{\EPIC}{\emph{EPIC}}
\newcommand{\SAS}{\emph{SAS}}
\newcommand{\rps}{\ensuremath{r_{\mathrm{ps}}}}
\newcommand{\rap}{\ensuremath{r_{\mathrm{ap}}}}
\newcommand{\rfhTX}{\ensuremath{r_{\mathrm{500}}(T_{\mathrm{X}})}}
\newcommand{\rfh}{\ensuremath{r_{\mathrm{500}}}}
\newcommand{\vel}{\ensuremath{v}}
\newcommand{\Ngal}{\ensuremath{N_{\mathrm{gal}}}}
\newcommand{\D}{\ensuremath{D}}
\newcommand{\dengal}{\ensuremath{\bar{\rho}_{\mathrm{gal}}}}
\newcommand{\pMpccu}{\ensuremath{\mbox{Mpc}^{-3}}}
\newcommand{\LB}{\ensuremath{L_{\mathrm{B}}}}
\newcommand{\Lsol}{\ensuremath{\mathrm{L_{\odot}}}}
\newcommand{\fsp}{\ensuremath{f_{\mathrm{sp}}}}
\newcommand{\LBGG}{\ensuremath{L_{\mathrm{BGG}}}}
\newcommand{\LXpLB}{\ensuremath{L_{\mathrm{X}}/L_{\mathrm{BGG}}}}
\newcommand{\dom}{\ensuremath{L_{\mathrm{12}}}}
\newcommand{\LXrfh}{\ensuremath{L_{\mathrm{X}}(r_{\mathrm{500}})}}
\newcommand{\ergpspLsol}{\ensuremath{\mbox{erg~s}^{-1}~\mathrm{L_{\odot}^{-1}}}}
\newcommand{\vgal}{\ensuremath{v_{\mathrm{gal}}}}
\newcommand{\vgroup}{\ensuremath{v_{\mathrm{group}}}}
\newcommand{\Hone}{\ensuremath{HI}}
\newcommand{\rchisq}{\ensuremath{\chi^2_\nu}}
\newcommand{\NH}{\ensuremath{N_{\mathrm{H}}}}
\newcommand{\pcmcu}{\ensuremath{\mbox{cm}^{-3}}}
\newcommand{\z}{\ensuremath{z}}
\newcommand{\Chandra}{\emph{Chandra}}
\newcommand{\kpc}{\ensuremath{\mbox{kpc}}}
\newcommand{\Tsh}{\ensuremath{T_{\mathrm{sh}}}}
\newcommand{\M}{\ensuremath{M}}
\newcommand{\vs}{\ensuremath{v_{\mathrm{s}}}}
\newcommand{\NASA}{\emph{NASA}}
\newcommand{\IPAC}{\emph{IPAC}}
\newcommand{\NED}{\emph{NED}}

\begin{document}

\maketitle

\label{firstpage}


\begin{abstract}

We present the results of a 24 \ks\ \XMM\ observation of the NGC\,5171 group of
galaxies.  NGC\,5171 is unusual in that it is an X-ray bright group (\LX\ $>$
10$^{42}$ \ergps), with irregular contours which are not centred on a bright
galaxy \citep[e.g.][]{mulchaey03}.  The global spectrum is adequately described
by a single temperature APEC model with \TX\ = 0.96$\pm$0.04 \kev, and \Z\ =
0.13$\pm$0.02 \Zsol, in good agreement with previous \ROSAT\ data.  We find the
X-ray contours are centred on a bright ridge of emission stretching from the BGG
to a nearby galaxy.  Spectral mapping reveals this ridge to be both cool (\TX\
$\approx$ 1.1 \kev) and metallic (\Z\ $\approx$ 0.4 \Zsol) in comparison to its
surrounding, suggesting it is the result of a tidal interaction between the two
galaxies.  Optical data reveals the member galaxies to have a high velocity
dispersion (\sigmav\ = 494$\pm$99 \kmps), and a significantly non-Gaussian
velocity distribution, suggesting the group is in the process of merging.  A
region of hot gas with \TX\ = 1.58$\pm$0.36 \kev\ is found to the West of the
bright central ridge, and we interpret this as shock-heating resulting from the
merging.  A further region of emission to the South-East of the bright central
ridge, with \TX\ = 1.14$\pm$0.13 \kev, is probably associated with a background
group, four times more distant.

\end{abstract}



\begin{figure*}
  \begin{minipage}[t]{241pt}
    \centering

    \includegraphics[width=0.9\linewidth]{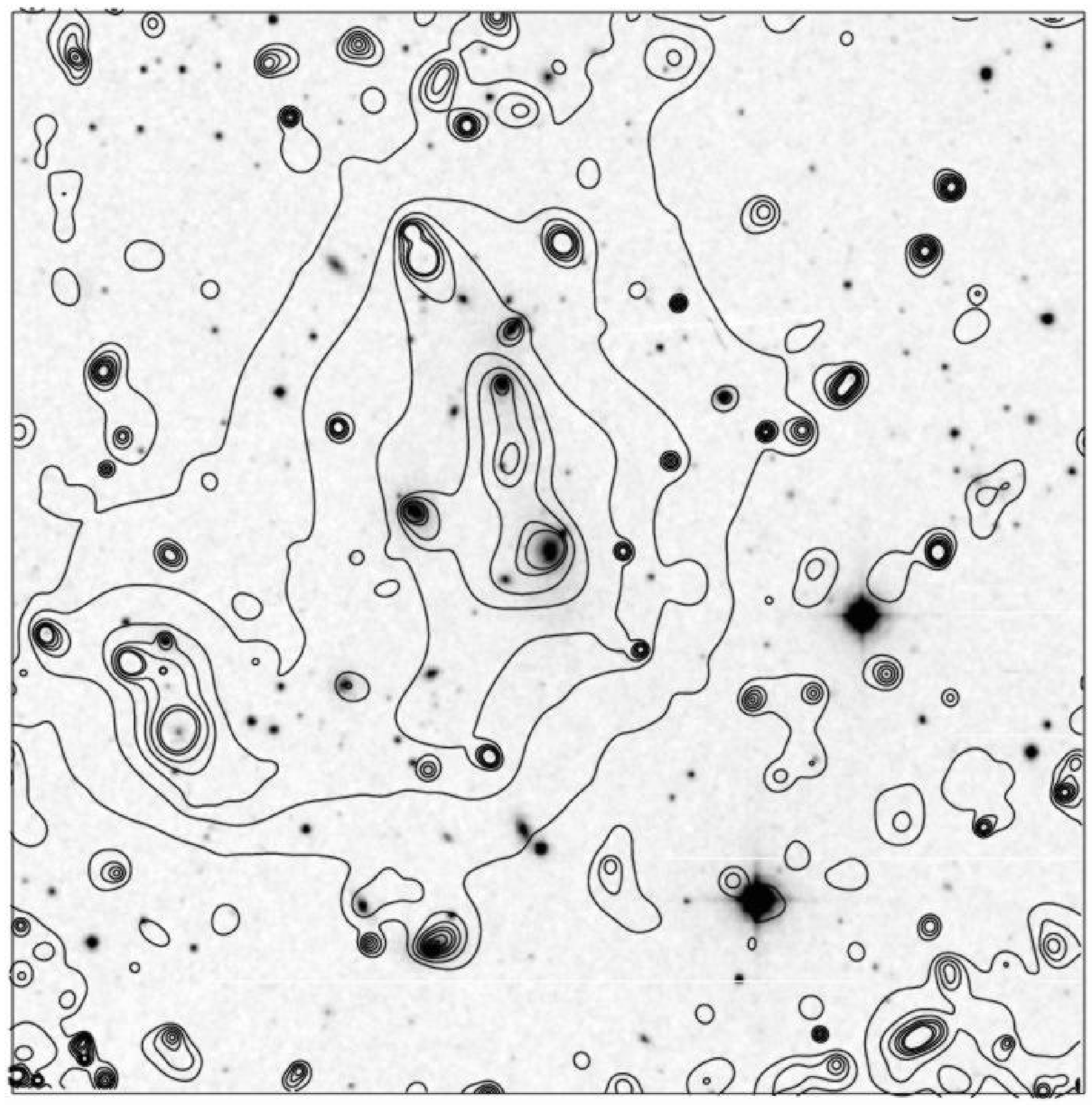}

  \end{minipage}\hspace{18pt}
  \begin{minipage}[t]{241pt}
    \centering

    \includegraphics[width=0.9\linewidth]{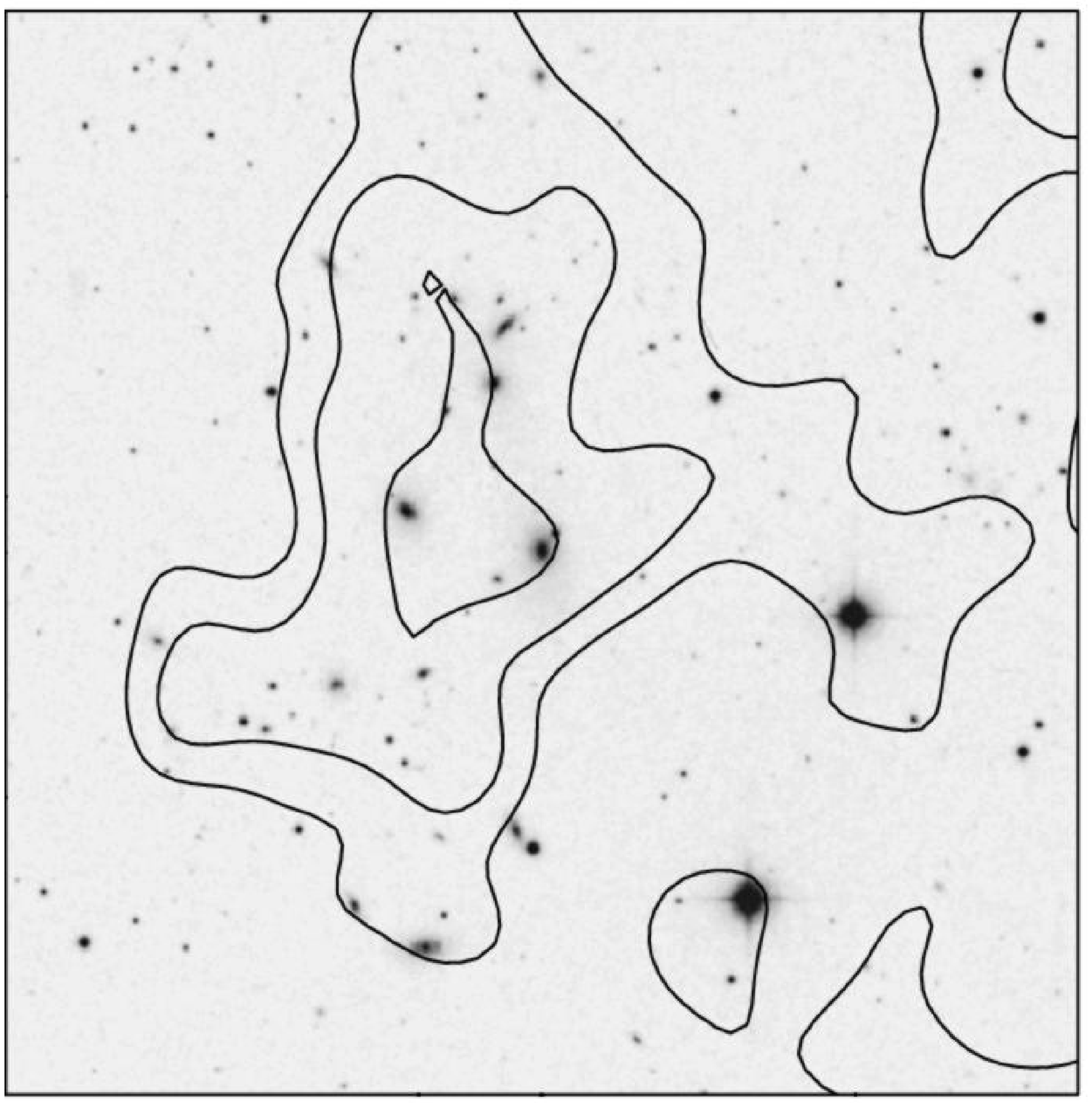}

  \end{minipage}\vspace{0.02\linewidth}
  \begin{minipage}{\linewidth}
    \centering

    \caption{X-ray/Optical overlays of the NGC 5171 group of galaxies.  Contours
             from the XMM mosaic (left) and the ROSAT PSPC (right) are overlaid
             on a B-band DSS image.  Each image is centred on the NGC\,5171
             galaxy, and covers the same region of sky as the X-ray images shown
             in Fig.~\ref{fig_mosaic}.}
    \label{fig_overlay}

  \end{minipage}
\end{figure*}


\section{Introduction}
\label{sec_intro}

The NGC\,5171 group of galaxies has been included in several large-sample studies
using the \ROSAT\ \PSPC\
\citep{helsdon00a,helsdon00b,mahdavi00,mulchaey03,osmond04a}.  Such studies have
evaluated the global properties of each group, and in many cases, used these
properties to parameterise mean X-ray/optical scaling relations.  However such
trends invariably include significant, non-statistical scatter in comparison to
the equivalent cluster trends \citep{helsdon00a}.  With the greater sensitivity
and energy resolution of \XMM\ we can corroborate the global values obtained
using \ROSAT, and map the spatial variations of these properties, in order to
investigate the origin of their scatter from mean trends.

X-ray imaging has shown the morphology of NGC\,5171 to be unusual in that despite
being X-ray luminous (\LX\ $>$ 10$^{42}$ \ergps), it exhibits irregular X-ray
contours which are not centred on a brightest group galaxy
(Fig.~\ref{fig_overlay}).  It is the only group in the \GEMS\ sample of 60 to
possess such properties.  Optical data reveals the member galaxies to have an
unusually high velocity dispersion for a poor group (\sigmav\ = 494$\pm$99 \kmps,
third highest in the \GEMS\ sample), and as such NGC\,5171 falls on the upper
envelope of the \sigmav-\TX\ relation. It has a value of \betaspec\ (the ratio of
specific energy in galaxies to that in the intragroup gas) of 1.43$\pm$0.59,
fourth highest in the \GEMS\ sample, whilst most groups have \betaspec\ less
than or equal to unity.

In this work we employ \XMM\ to map the spectral properties of NGC\,5171.  We
then combine our X-ray results with a study of the dynamics of the optical
galaxies, in order to better understand the evolutionary processes responsible
for the unusual properties of this group.

Where applicable we have compared our results to those of the Group Evolution
Multi-wavelength Study \citep[\GEMS,][]{osmond04a}, who included a 5 ks
observation of NGC\,5171, in their \ROSAT-\PSPC\ study of a sample of 60
optically selected groups.  This comparison provides some indication of the
reliability of the \ROSAT\ data, as well as demonstrating the additional power of
\XMM.

This paper is organised as follows: In $\S$ \ref{sec_reduction} we describe the
X-ray data reduction, and in $\S$ \ref{sec_spatial} we present the results of the
subsequent spatial analysis.  $\S$ \ref{sec_optical} includes an explanation of
the optical data relating the member galaxies.  In $\S$ \ref{sec_spectral} we
describe the global spectral analysis, and in $\S$ \ref{sec_specmap} we describe
the spatial variation of these spectral properties.  In $\S$ \ref{sec_discussion}
we discuss the implications of the X-ray and optical results, and in $\S$
\ref{sec_conclusions} we summarise our conclusions.

Throughout this work we use \Hzero\ = 70 \kmpspMpc, and a distance to NGC\,5171
of 107 \Mpc, based on an average group redshift, corrected for infall into Virgo
and the Great Attractor.


\section{Data Reduction}
\label{sec_reduction}

NGC\,5171 was observed for 24 ks with \XMM\ on 30 December 2001
(Obs. Id. 0377-0041180801).  Data from the European Photon Imaging Camera (\EPIC)
were prepared for examination using the EPCHAIN and EMCHAIN pipeline processing
routines within the Science Analysis Software (\SAS).  Periods of significant
X-ray flaring were identified using a binned light-curve.  Times for which the
X-ray flux exceeded the mean by $\geq$ 3$\sigma$ were removed, leaving 19 ks of
usable data.

PN events with pattern $>$ 4 and MOS events with pattern $>$ 12 were removed
from the data, and further bad events were excluded using the XMMEA event
attribute filters.  Point sources were identified from data within 5 energy bands
(0.2-0.5, 0.5-2.0, 2.0-4.5, 4.5-7.5, 7.5-12.0 \kev) using both a sliding box
algorithm (EBOXDETECT) and a maximum likelihood point spread function fitting
(EMLDETECT).  Results of the two methods were combined to give 107 detected point
sources in the field of view.  Five of these sources formed a chain between two
central galaxies, and due to their arrangement, were unlikely to represent
genuine point sources.  We conclude that a bright filament at the centre of the
group (Section \ref{sec_spatial}) has confused the point source searching, and we
have therefore not excluded this region from our subsequent analysis of the
diffuse X-ray emission.  The remaining 101 sources, including all those
coincident with galaxies, were excluded to the 80\% radius for 5 $\kev$ photons
(\rps\ = 25 arcsec).

Background properties were estimated from a superposition of 72 observations of a
blank region of sky taken with the same instrument, mode and filter as the
NGC\,5171 observation \citep{read03}.  Background data were filtered in the same
way as the principal observation, and residuals in the background emission,
between the source and blank-sky datasets, were estimated from an annulus at the
edge of the field of view (12 $<$ $r$ $<$ 16 arcmin).  Such residuals may arise
from celestial variations in galactic X-ray emission, and soft proton emission
\citep{lumb02}, or from secular evolution of the instrumental background
\citep{deluca04}.  Spectra were extracted for all three instruments and fitted
over the range 0.3 - 12 \kev, excluding 7.0 - 9.0 \kev.  The model used included
a diffuse gas component \citep[APEC,][]{smith01} to account for variations in the
galactic X-ray emission, and a soft power-law component to account for variations
in soft proton emission.  The PN instrument required an additional hard power-law
component (index $\sim$ 0) to account for secular evolution of the background.
The abundance and redshift of the galactic component were frozen at 1 and 0
respectively, and the power-law components were allowed to vary between
instruments.  A radial profile of surface brightness confirmed that the annulus
used to evaluate the background residuals contained essentially no group emission
(Section \ref{sec_spatial}).  Furthermore, a second APEC component added to the
background residual model, achieved no significant improvement in the quality of
the fit.  For an account of a similar background subtraction approach see
\citet{zhang04}.  Results of the fit to the residual background are presented in
Table \ref{tab_spectral}.

An energy range of 0.4 - 2.0 $\kev$ was adopted, as the range in which group
X-ray emission dominated over the background, and data outside this range were
discarded when generating images and spectra.

\begin{figure*}
   \centering

  \includegraphics[width=0.7\linewidth]{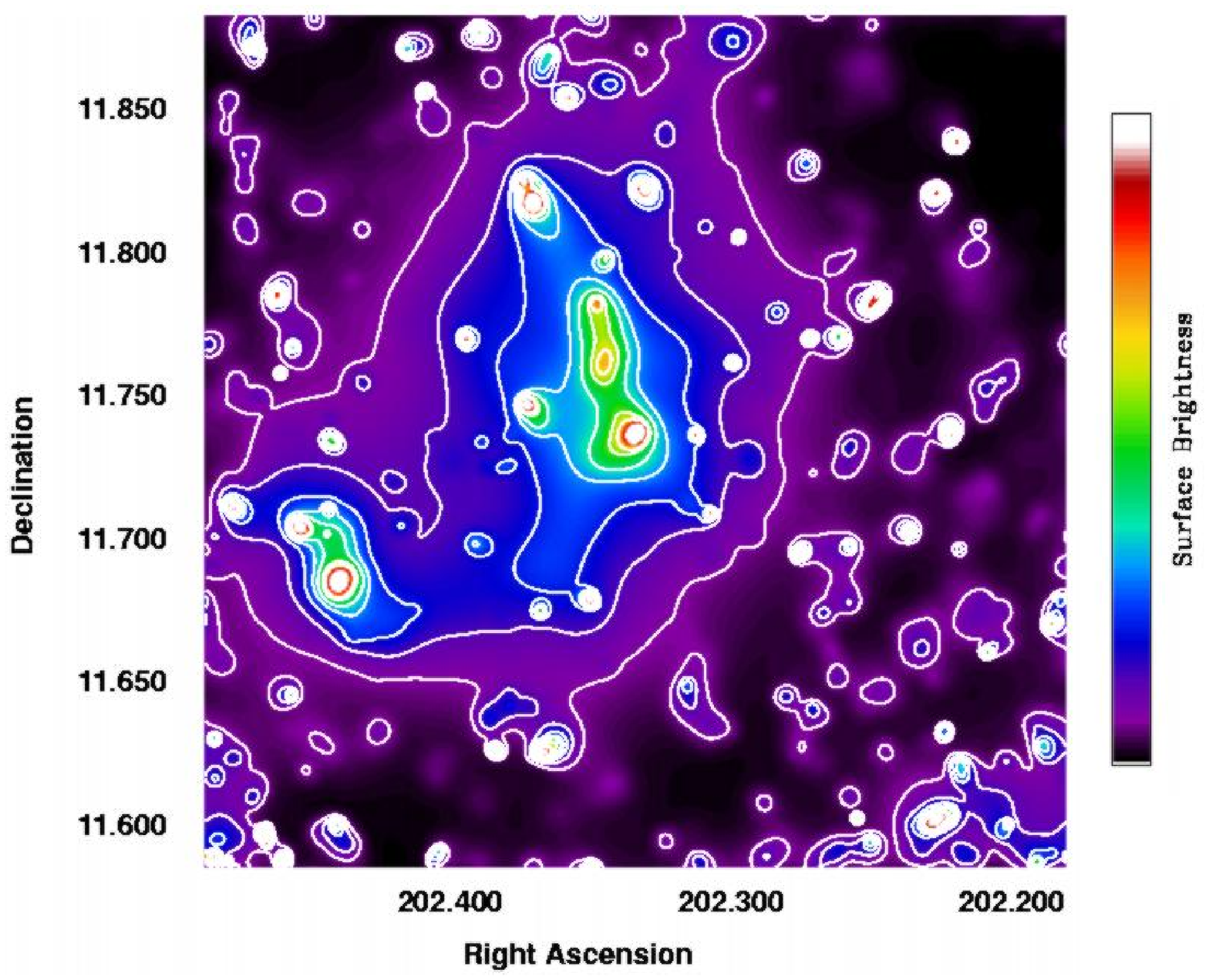}
  \vspace{5pt}\\
  \includegraphics[width=0.7\linewidth]{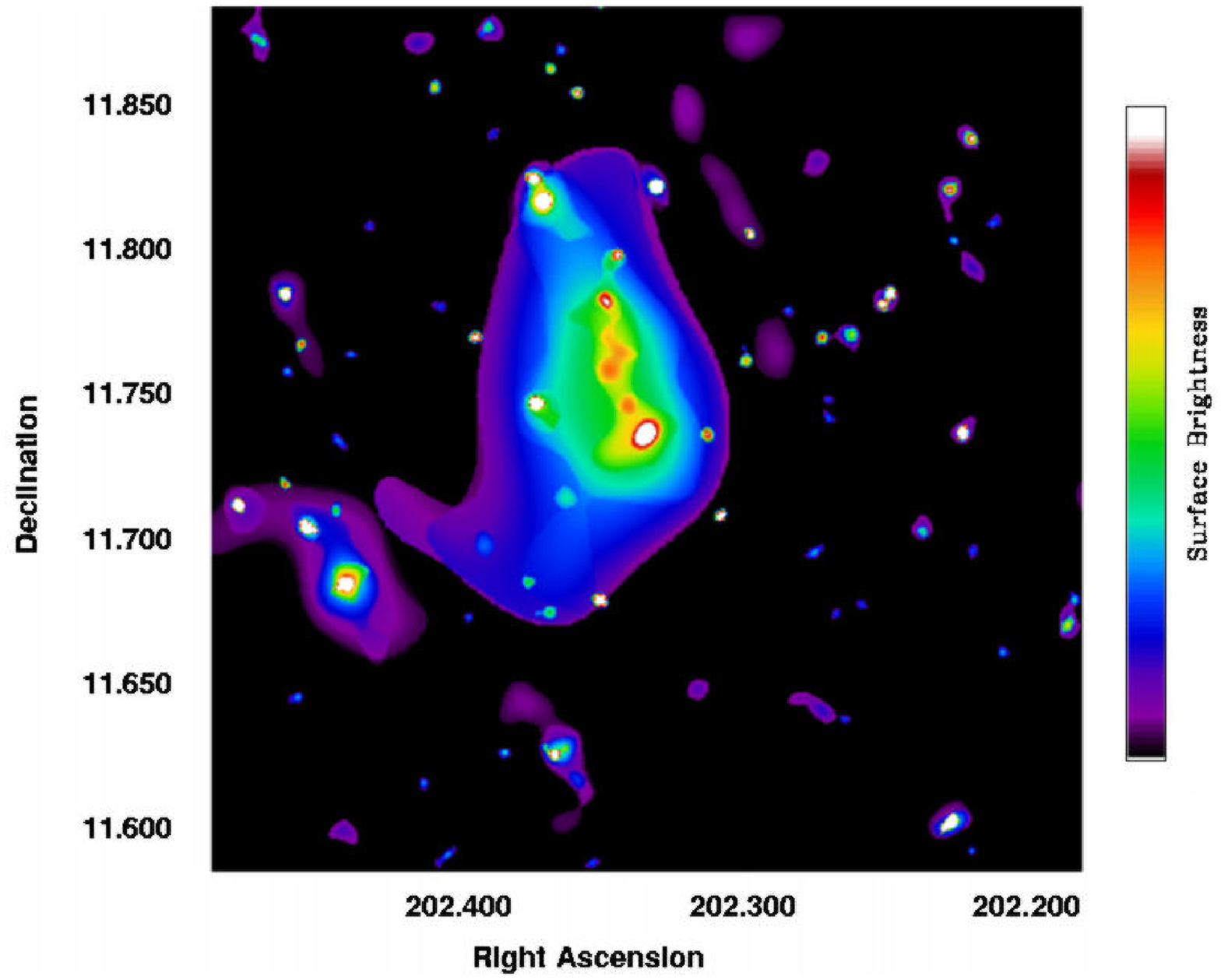}

  \caption{Mosaiced images of NGC 5171.  Data from PN, MOS1 and MOS2 were
           combined and smoothed using an adaptive smoothing algorithm (top) and
           a wavelet smoothing algorithm (bottom).  Contours in the top figure
           represent logarithmically spaced surface-brightness isophotes, and are
           included for comparison with other images.}
  \label{fig_mosaic}

\end{figure*}


\section{Spatial Analysis}
\label{sec_spatial}

Images were created by generating cleaned events lists for each instrument, from
within a square region centred on the brightest group galaxy (BGG) NGC\,5171, and
binning into 1 arcsec pixels.  Particle flux was estimated from a dataset taken
with the filter wheel set to closed \citep{marty02}, and normalised using the
counts in the range 10 - 12 \kev, in which the emission was assumed to be entirely
particle in origin.  This particle estimate was then subtracted from the image
before the exposure correction was applied, as the particle flux is not focused
by the X-ray mirrors.  An X-ray background map was estimated from a blank-sky
image processed in the same way, and then subtracted from the source image.
Images from the 3 instruments were combined to form a mosaic, and smoothed using
an adaptive smoothing algorithm.  The smoothing scale was varied so as to
maintain a $>$ 3$\sigma$ significance in surface brightness, using a background
level estimated from an annulus surrounding the kernel.  For comparison purposes,
the mosaic was also smoothed using a wavelet smoothing algorithm
\citep{vikhlinin98}, where the smoothing scale was varied between 2 and 7 arcsec,
and regions of $>$ 4$\sigma$ significance from each scale co-added.
Fig.~\ref{fig_mosaic} shows the results of the two methods, with logarithmically
spaced surface-brightness contours overlaid on the adaptively smoothed image.  We
can see from these figures that the two techniques produce similar results.
However it appears that the wavelet smoothing has failed to recover the extended,
low surface-brightness emission seen in the adaptively smoothed image.  A radial
profile extracted from an unsmoothed image reveals genuine emission out to the
edge of the image, suggesting that the wavelet smoothing algorithm is less
effective at reduced intensity.

Fig.~\ref{fig_overlay} shows logarithmically spaced surface-brightness contours
taken from the adaptively smoothed mosaic, overlaid on a B-band DSS image, and
for comparison the equivalent image generated from a 5 ks \ROSAT\ \PSPC\
pointing.  These images show that the group-scale X-ray emission previously
detected with \ROSAT\ has been resolved into two distinct sub-regions.  The
larger of the two sub-regions exhibits X-ray contours which are highly elongated
in approximately the North-South direction, suggesting that the corresponding gas
is not relaxed.  Furthermore, a bright filament of emission (hereafter the bright
central filament) can be seen at the centre of this sub-region, orientated in a
similar North-South direction.  This filament appears to join two galaxies: the
brightest group galaxy NGC\,5171, at the Southern end, and NGC\,5176 at the
Northern end.  In contrast the South-East emission region has only a faint
central galaxy.  Finally this comparison reveals that the Western extension
apparent in the \ROSAT\ image is not present in the \XMM\ image, suggesting it
may be the result of point-source confusion in the inferior \ROSAT\ data.

A radial surface-brightness profile centred on NGC\,5171 was extracted from an
unsmoothed mosaic of images from all three instruments, with point sources
removed to \rps\ (25 arcsec), and the South-East emission region removed to 2
arcmin.  The radius at which the surface brightness dropped to approximately the
background level was used to define an aperture radius of \rap\ = 9 arcmin.  This
is consistent with the adaptively smoothed mosaic, which demonstrates emission
dropping to approximately the background level (black) at the edge of the image.
The radial profile also confirms that the region used to evaluate background
residuals is largely free of emission from NGC\,5171.

All images presented here have a width equivalent to twice the aperture radius.


\section{Optical Membership}
\label{sec_optical}

We derive our optical galaxy membership using the procedure of \citet{osmond04a}.
An overdensity radius is defined from the group X-ray temperature using a
relation derived from simulations \citep{evrard96}:

\begin{equation}
\rfhTX ~ = ~ \frac{124}{\Hzero} \sqrt{\frac{\TX}{10~\kev}} ~ \Mpc,
\label{eqn_r500_TX}
\end{equation}

\noindent Galaxies are then extracted from the \NASA-\IPAC\ Extragalactic
Database (\NED) within a radius \rfh\ of the BGG position, and 3 standard
deviations (\sigmav) of the group velocity (\vel), and the process iterated using
the recalculated values of \vel\ and \sigmav.  NGC\,5171 is a typical system in
comparison to the remainder of the \GEMS\ sample of 60 groups, in all parameters
except \sigmav\ where it has the third highest value (494 \kmps), and spiral
fraction, where it has the joint lowest (0).  Table \ref{tab_optical} summarises
the optical properties of NGC\,5171.

\begin{table}
\begin{center}
\small
\begin{tabular}{@{}llr@{}}
\hline


\multicolumn{2}{@{}l}{Parameter}     &  Value           \\

\hline

\Ngal      &                      &  12              \\
\vel       &  (\kmps)             &  6924$\pm$132    \\
\sigmav    &  (\kmps)             &  494$\pm$99      \\
\D         &  (\Mpc)              &  107             \\
\rfh       &  (\Mpc)              &  0.58            \\
\dengal    &  (\pMpccu)           &  15              \\
\LB        &  (log \Lsol)         &  11.28           \\
\fsp       &                      &  0.00            \\
\LBGG      &  (log \Lsol)         &  10.76           \\
\dom       &                      &  2.65            \\

           &                      &                  \\

\TX        &  (\kev)              &  1.07$\pm$0.09   \\
\Z         &  (\Zsol)             &  1.47$\pm$1.25   \\
\LX        &  (log \ergps)        &  42.38$\pm$0.06  \\
\LXrfh     &  (log \ergps)        &  42.45$\pm$0.06  \\

           &                      &                  \\

\betaspec  &                      &  1.43$\pm$0.59   \\
\LXpLB     &  (log \ergpspLsol)   &  31.11$\pm$0.06  \\


\hline
\end{tabular}
\end{center}


\caption
{\label{tab_optical}  A summary of the X-ray and optical data for NGC\,5171, as
                      derived by \citet{osmond04a}}


\end{table}

\begin{figure}
  \centering
 
  \includegraphics[width=\linewidth]{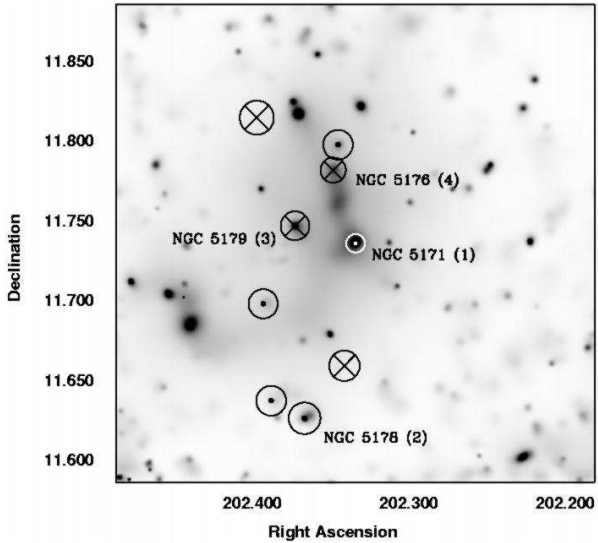}
  \caption{Galaxy velocity map.  Each symbol represents a group galaxy where the
           size of the symbol is proportional the galaxy velocity in the
           rest-frame of the group $(|\vgal-\vgroup|)$.  The symbol $\otimes$
           represents galaxies moving away, and $\odot$ galaxies moving towards
           the observer relative to the mean group velocity.  The four brightest
           group galaxies are labelled with their name and brightness rank, and
           the underlying grey scale image is the adaptively smoothed mosaic
           shown in Fig.~\ref{fig_mosaic}.}
  \label{fig_galmap}

\end{figure}

To better illustrate the spatial and velocity distribution of the member galaxies,
we have overlaid a galaxy velocity map on an image of the X-ray emission
(Fig.~\ref{fig_galmap}).  Each symbol represents a group galaxy, where the size of
the symbol is proportional the line-of-sight galaxy velocity in the rest-frame of
the group $(|\vgal-\vgroup|)$.  The symbol $\otimes$ represents galaxies moving
away, and $\odot$ galaxies moving towards the observer.  The three group member
galaxies not included in Fig.~\ref{fig_galmap} are located at a similar RA to the
BGG (NGC\,5171), but lie 3 arcmin outside the X-ray field of view, to the North.

One significant feature of this figure is that none of the member galaxies of
NGC\,5171 coincide with the South-East emission region.  A search of \NED\ within
2 arcmin of the position of this region reveals 3 galaxies with undetermined
redshifts.  However if these galaxies lie at approximately the same redshift as
our group then one would expect their redshifts to have been determined and
included in \NED.  Furthermore, the average apparent luminosity of these galaxies
is $\approx$ 16 times fainter than the average for the member galaxies of
NGC\,5171, suggesting they may be $\approx$ 4 times more distant.  It is
therefore likely that these galaxies, and their corresponding X-ray emission, are
associated with a more distant group.  As such we have excluded this region from
the global spectral analysis.

\section{Spectral Analysis}
\label{sec_spectral}

For each instrument, a source spectrum was created from cleaned events within a
radius r $<$ \rap\ of the group position, and a background spectrum created from
the same region of the blank-sky dataset.  Energy channels were grouped into bins
containing $\geq$ 30 counts, and a Redistribution Matrix File (RMF) and Ancillary
Response File (ARF) generated using the relevant SAS commands.  The three spectra
were then simultaneously fitted with a single-component diffuse gas emission
model (APEC), and a multiplicative absorption component (WABS), using XSPEC
\citep{arnaud96}.  All fits were performed with the neutral hydrogen column
density fixed at a value taken from \Hone\ radio observations \citep{dickey90},
and with the redshift fixed at the group value derived by \citet{osmond04a}.
Residual background, following the subtraction of the blank-sky background, was
accounted for by including the residual model described in Section
\ref{sec_reduction} in the spectral fitting.  As variations in this residual
background across the detector are not well understood, the normalisations of the
residual background model components were allowed to vary.

An inspection of the residuals from the fitted model revealed two additional
emission lines in the MOS spectra, which were not accounted for by the model.
MOS data were refitted with a variable abundance model (VAPEC), with abundances
taken from \citet{anders89}, but none of the additional lines accounted for
these features.  However adding two Gaussian components to the existing model,
and fitting simultaneously to only the MOS spectra, identified the energies of
these lines as 1.50 and 1.76 \kev\ respectively, and improved the fit
significantly ($\Delta$\rchisq\ = 0.51).  These energies correspond to the Al-K
and Si-K instrumental fluorescence lines, originating from the camera shielding
and CCD substrate respectively \citep{lumb02}.  To exclude these lines we have
ignored all MOS data above 1.4 \kev\ in the spectral fitting.  Data were refitted
with a variable abundance model (VAPEC), a two temperature model (APEC+APEC), and
a differential emission model (CEMEKL), but no significant improvement was
obtained.

\begin{figure}
  \centering

  \includegraphics[height=\linewidth,angle=270]{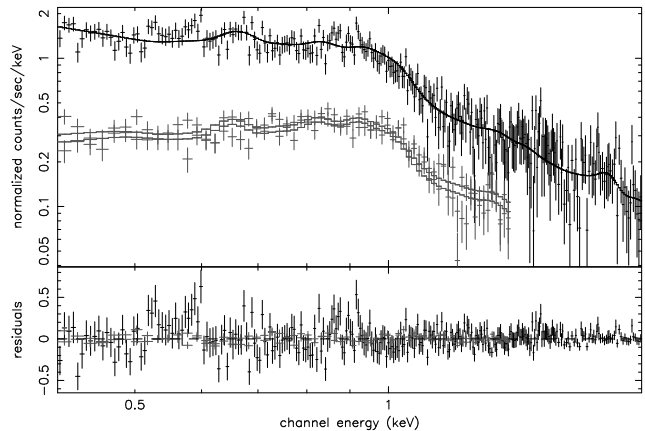}
  \caption{Spectral data from all three EPIC instruments, shown with the
           best-fitting APEC model, and the residuals from this model.  The upper
           model corresponds to PN data, and the lower two models to MOS data.}
  \label{fig_spectrum}

\end{figure}

Results of the spectral fitting are presented in Table \ref{tab_spectral} and the
spectral data, together with the best-fitting APEC model, and the residuals from
this model, are plotted in Fig.~\ref{fig_spectrum}.

\begin{table}
\begin{center}
\small
\begin{tabular}{@{}lcllr@{}}
\hline


Comp.     &  Ins.  & \multicolumn{2}{l}{Parameter}     &  Value           \\

\hline             

IGM       &        & \TX         &  (\kev)              &  0.96$\pm$0.04   \\
          &        & \Z          &  (\Zsol)             &  0.13$\pm$0.02   \\
          &        & \LX         &  (log \ergps)        &  42.54$\pm$0.08  \\

          &        & \NH         &  (10$^{22}$ \pcmcu)  &  0.0193          \\
          &        & \z          &                      &  0.0230          \\

          &        & \betaspec   &                      &  1.59$\pm$0.46   \\
          &        & \LXpLB      &  (log \ergpspLsol)   &  31.26$\pm$0.08  \\

          &        & \rchisq\    &                      &  1.21            \\

\hline             

Galactic  &        & \TX\        &  (\kev)              &  0.31$\pm$0.02   \\

\hline             

Soft      &  PN    & Index       &                      &  2.93$\pm$0.29   \\
                   &  M1    & Index       &                      &  1.81$\pm$0.89   \\
                   &  M2    & Index       &                      &  1.52$\pm$0.40   \\

\hline             

Secular   &  PN    & Index       &                      &  0.08$\pm$0.07   \\


\hline
\end{tabular}
\end{center}


\caption
{\label{tab_spectral}  The spectral properties of NGC 5171 and the associated
background.  The latter 3 components describe the residual emission following
the blank-sky subtraction (Section \ref{sec_reduction}), and correspond to
variations in the galactic emission, soft proton emission and secular evolution
of the PN instrument respectively.}


\end{table}

\section{Spectral Mapping}
\label{sec_specmap}

In order to better understand the spatial distribution of the spectral properties
of NGC\,5171, we have derived spectral maps using two complementary methods:

\subsection{Adaptively Binned Spectral Maps}
\label{sec_specmap_adapt}

Data were extracted from within a square region centred on the BGG, with a
width of 2\rap\ (18 arcmin), and background data were extracted from the same
region of the blank-sky dataset.  The square was subdivided into
2$^{n}\times$2$^{n}$ bins, where n = 1 initially, and any bins containing a total
of $\geq$ 500 counts over all three instruments after background subtraction,
were used to generate spectra.  RMF and ARF files were generated corresponding
to 6 annular regions of equal width, centred on the bore-sight.  The spectra
for each bin were then associated with the most appropriate RMF and ARF, and
simultaneously fitted with an absorbed APEC model using XSPEC.  Residual
background following the subtraction of the blank-sky background, was accounted
for as described in Section \ref{sec_spectral}.  The process was repeated with n
increasing by 1 on each iteration, until it produced a layer with no bins
containing $\geq$ 500 counts.  Any bins containing $<$ 500 counts adopted the
properties of the corresponding larger region from the previous layer.

\begin{figure}
  \centering

  \includegraphics[width=\linewidth]{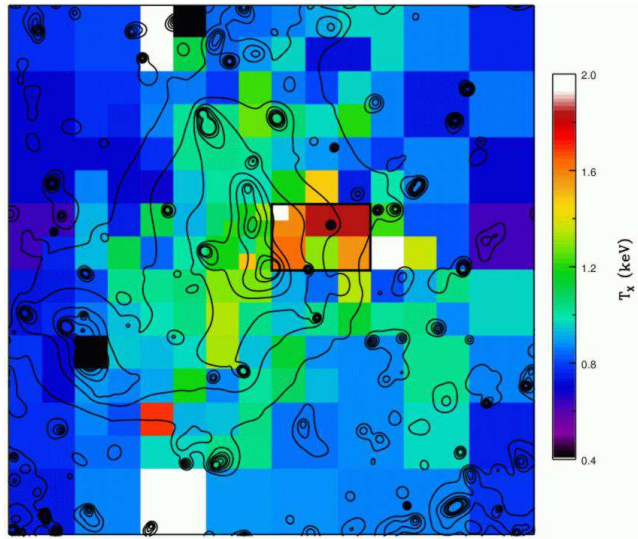}
  \caption{Adaptively binned temperature map of NGC\,5171.  The contours are
             the logarithmically spaced surface brightness isophotes shown in
             Fig.~\ref{fig_mosaic}, and the rectangle represents the spectral
             extraction region for the Western Hot Region}
  \label{fig_specmap_adapt}

\end{figure}

Fig.~\ref{fig_specmap_adapt} shows the temperature map derived using this method.
We can see that the IGM of NGC\,5171 exhibits temperatures of $\approx$ 0.8 \kev\
in the outer regions with a rise to $\approx$ 1.3 \kev\ in the centre.  The bright
central filament has a temperature which is generally consistent with its
surroundings.  There is also evidence for a region of hot gas to the West of the
bright central filament (hereafter the Western hot region), with a temperature
of $\approx$ 1.7 \kev, but with no corresponding increase in surface-brightness.
A metallicity map derived in the same way shows the majority of the group gas to
have a metallicity of \Z\ $\sim$ 0.1 - 0.4 \Zsol.  The metal abundance in the
Western hot region is $\approx$ 0.4 \Zsol.

\subsection{Surface Brightness Region Mapping}
\label{sec_specmap_region}

\begin{figure*}
  \begin{minipage}[t]{241pt}
    \centering

    \includegraphics[width=0.9\linewidth]{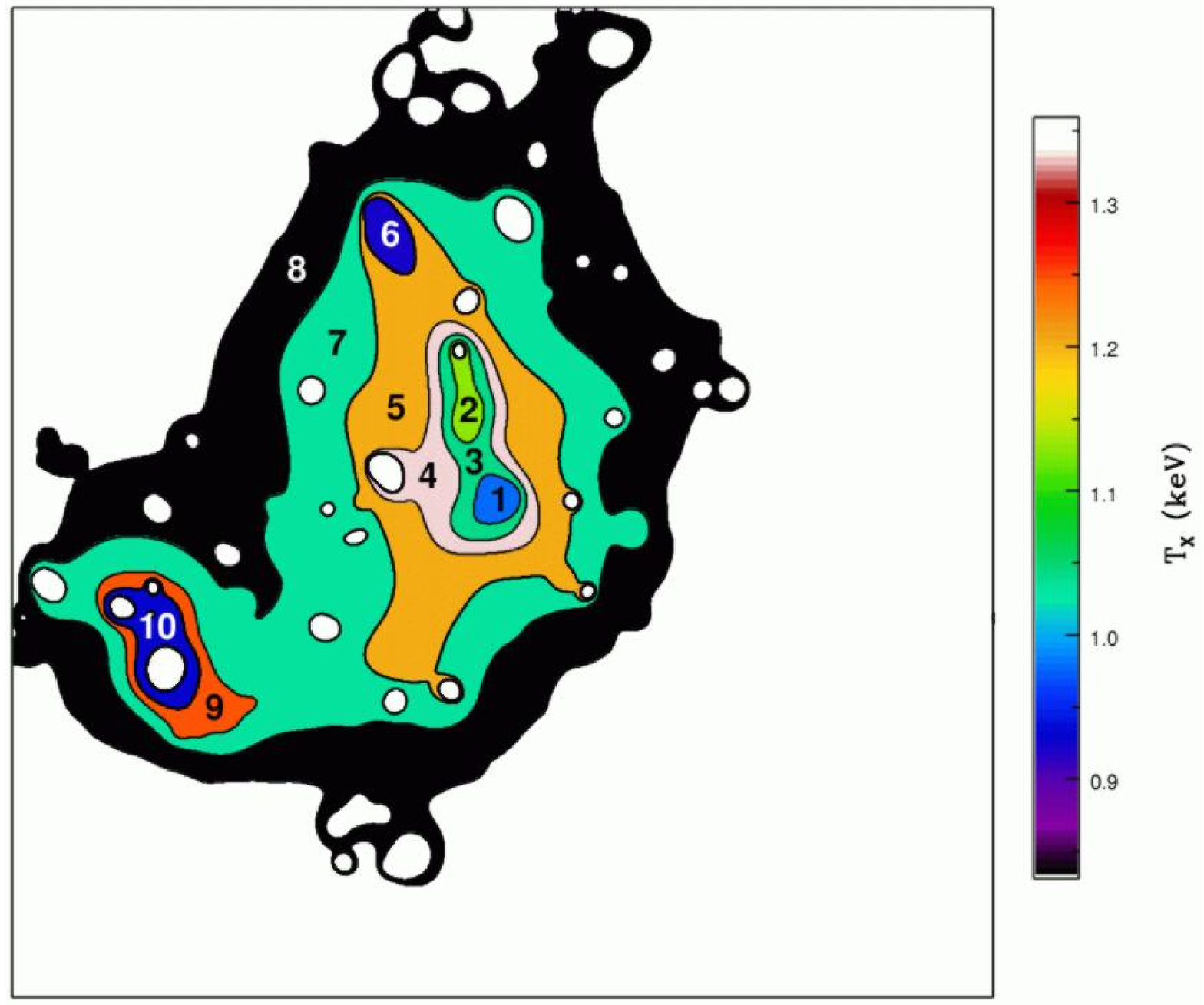}

  \end{minipage}\hspace{18pt}
  \begin{minipage}[t]{241pt}
    \centering

    \includegraphics[width=0.9\linewidth]{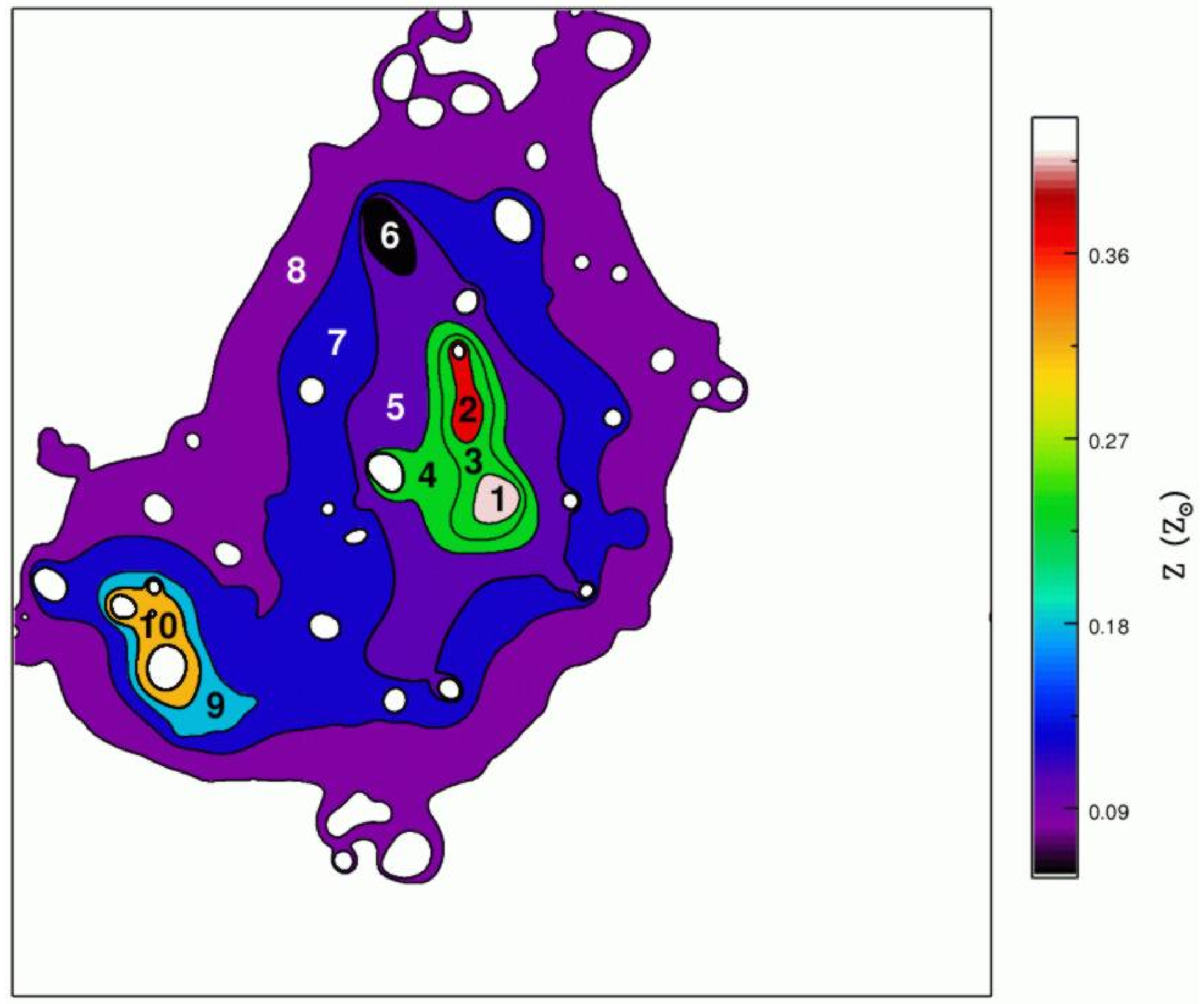}

  \end{minipage}\\
  \begin{minipage}{\linewidth}
    \centering

    \caption{Spectral maps showing the temperature (left) and metallicity (right)
             of NGC\,5171 in regions of similar surface brightness. The contours
             are the logarithmically spaced surface brightness isophotes shown in
             Fig.~\ref{fig_mosaic}.  The parameters and errors corresponding to
             each region are shown in Table~\ref{tab_region}.}
    \label{fig_specmap_region}

  \end{minipage}
\end{figure*}

Regions of similar surface brightness were defined from an adaptively smoothed
mosaic image, using thresholds equal to the contour levels shown in
Fig.~\ref{fig_mosaic}.  Regions containing $>$ 200 counts above the background,
across all three instruments, were fitted with an absorbed APEC model.
Fig.~\ref{fig_specmap_region} shows the temperature and metallicity maps derived
using this method, and Table~\ref{tab_spectral} details the corresponding
parameter values and errors.  In agreement with Fig.~\ref{fig_specmap_adapt}, the
temperature map shows a general rise in temperature from $\approx$ 0.8 \kev\ in
the outskirts (region 1) to $\approx$ 1.3 \kev\ in the centre (region 4).  The
Western hot region seen in Fig.~\ref{fig_specmap_region} exhibits no
corresponding increase in surface brightness, and as such does not feature in
this temperature map.  However we do see a drop in temperature to $\approx$ 1.1
\kev\ in the bright central filament, which is not apparent in
Fig.~\ref{fig_specmap_adapt} due to the size and position of the bins.  Region 1
shows a further dip in temperature to $\approx$ 1.0 \kev, corresponding to
emission from the BGG.  The metallicity map shows a reasonably flat distribution
at \Z\ $\approx$ 0.10 \Zsol\ across most of the group, with a rise to \Z\ $>$
0.20 \Zsol\ in the bright central filament.  Such a gradient in metal abundance
suggests enrichment of the IGM by supernovae in the central galaxies
\citep{buote03}.

\begin{table}
\begin{center}
\small
\begin{tabular}{@{}lccr@{}}
\hline


Region  &  \TX     &  \Z       & \rchisq     \\
        &  (\kev)  &  (\Zsol)  &             \\

\hline

1       &  0.98$\pm$0.14  &  0.41$\pm$2.44  & 1.37 \\
2       &  1.13$\pm$0.18  &  0.37$\pm$1.10  & 2.65 \\
3       &  1.04$\pm$0.16  &  0.23$\pm$0.20  & 0.79 \\
4       &  1.33$\pm$0.15  &  0.23$\pm$0.27  & 0.87 \\
5       &  1.20$\pm$0.14  &  0.10$\pm$0.07  & 0.98 \\
6       &  0.92$\pm$0.24  &  0.06$\pm$0.16  & 0.73 \\
7       &  1.03$\pm$0.12  &  0.12$\pm$0.08  & 1.15 \\
8       &  0.83$\pm$0.10  &  0.08$\pm$0.06  & 0.87 \\
9       &  1.25$\pm$0.41  &  0.18$\pm$0.69  & 1.04 \\
10      &  0.92$\pm$0.15  &  0.32$\pm$2.48  & 0.77 \\


\hline
\end{tabular}
\end{center}


\caption
{\label{tab_region}  The spectral properties of NGC\,5171 in regions of similar
surface brightness, as defined by the contours shown in Fig~\ref{fig_mosaic}.
Values correspond to the regions marked in Figure \ref{fig_specmap_region}.}


\end{table}



\section{Discussion}
\label{sec_discussion}

\subsection{\ROSAT\ comparison}

Table~\ref{tab_optical} details the X-ray properties of NGC\,5171 derived using
the \ROSAT\ \PSPC.  The \ROSAT\ values of \LX\ = 42.38$\pm$0.06 \ergps, and \TX\
= 1.07$\pm$0.09 \kev, are consistent with the values derived here ($\Delta$\LX\ =
1.6$\sigma$, $\Delta$\TX\ = 1.1$\sigma$).  Furthermore, the metallicity previously
unconstrained by \ROSAT\ is now well defined.

This comparison provides some evidence that \ROSAT\ \PSPC\ can be used to evaluate
reliable global properties of groups, despite lacking the energy resolution to
determine metallicity.  Furthermore, our comparison corresponds to a group in the
process of merging, and one might expect better agreement for more relaxed systems.
\citet{helsdon04a} compare \ROSAT\ and \Chandra\ results for two low velocity
dispersion groups and find that, even in these low luminosity cases, the diffuse
X-ray emission identified by \ROSAT\ is not grossly misleading.  However point
source contamination and inaccurate spectral characterisation lead to an
overestimation of \LX\ by $\sim$ 30 - 40\%, although source contamination should
be less of a problem for more luminous systems (e.g. \LX\ $>$ 10$^{41}$ \ergps).
We can therefore have some confidence in the scaling relations derived using
\ROSAT\ data.  However, the superior sensitivity of \XMM, and in this case longer
exposure time, reveals additional features in the morphology, and allows us to
map the spatial variation of spectral properties.

\subsection{Scaling relations}

NGC\,5171 exhibits an unusually high velocity dispersion (\sigmav\ = 494$\pm$94
\kmps) in comparison to typical groups in the sample from which it was taken.
However the presence of a hot, intragroup medium detectable out to $\approx$ 300
\kpc\ confirms that it is a gravitationally bound group rather than a chance
alignment of galaxies.  The high value of \betaspec\ = 1.59 indicates that there
is significantly more specific energy in the galaxies than in the gas, in
contrast to most groups \citep{helsdon00a,osmond04a}.

\begin{figure*}
  \centering

  \includegraphics[angle=270,width=0.49\linewidth]{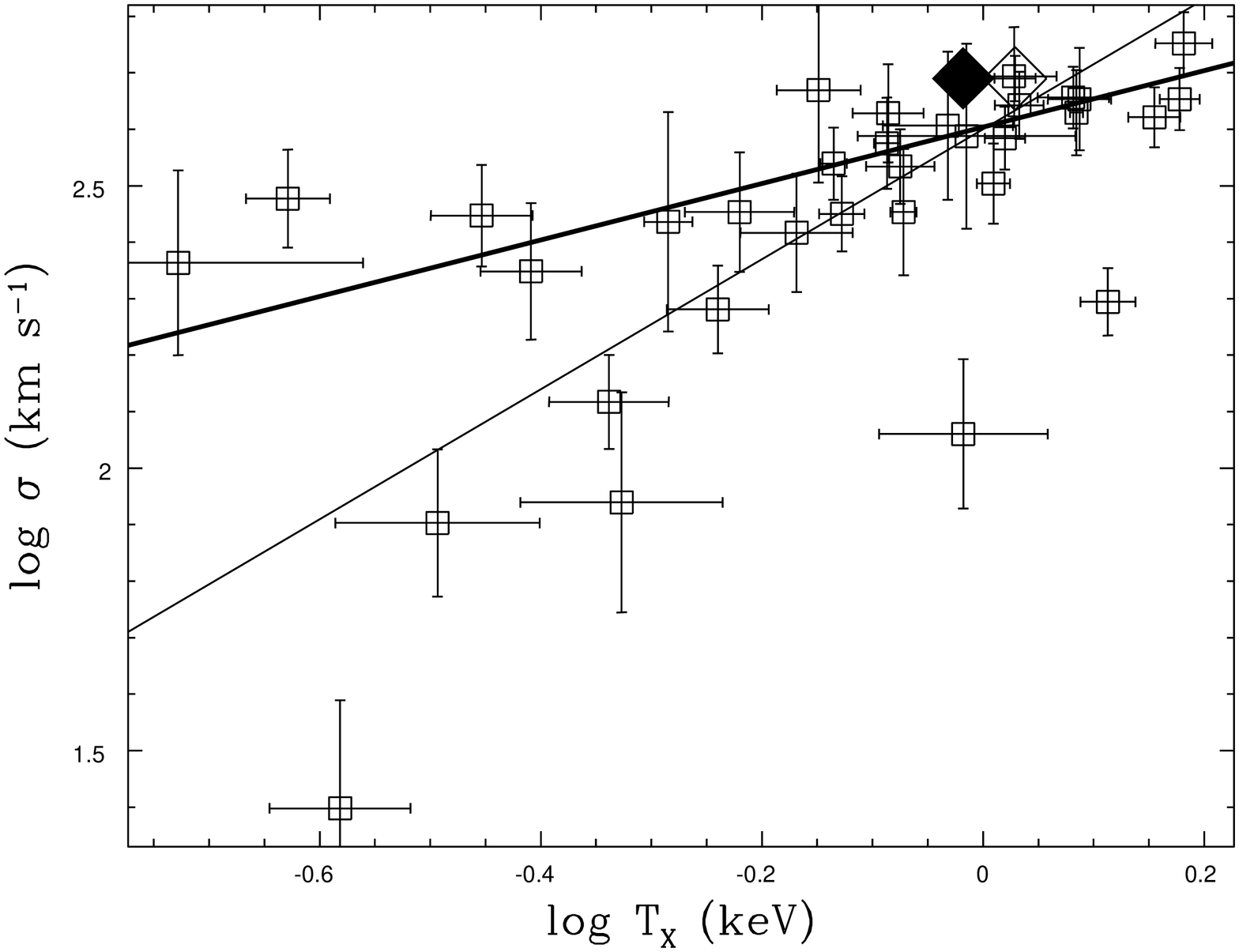}
  \includegraphics[angle=270,width=0.49\linewidth]{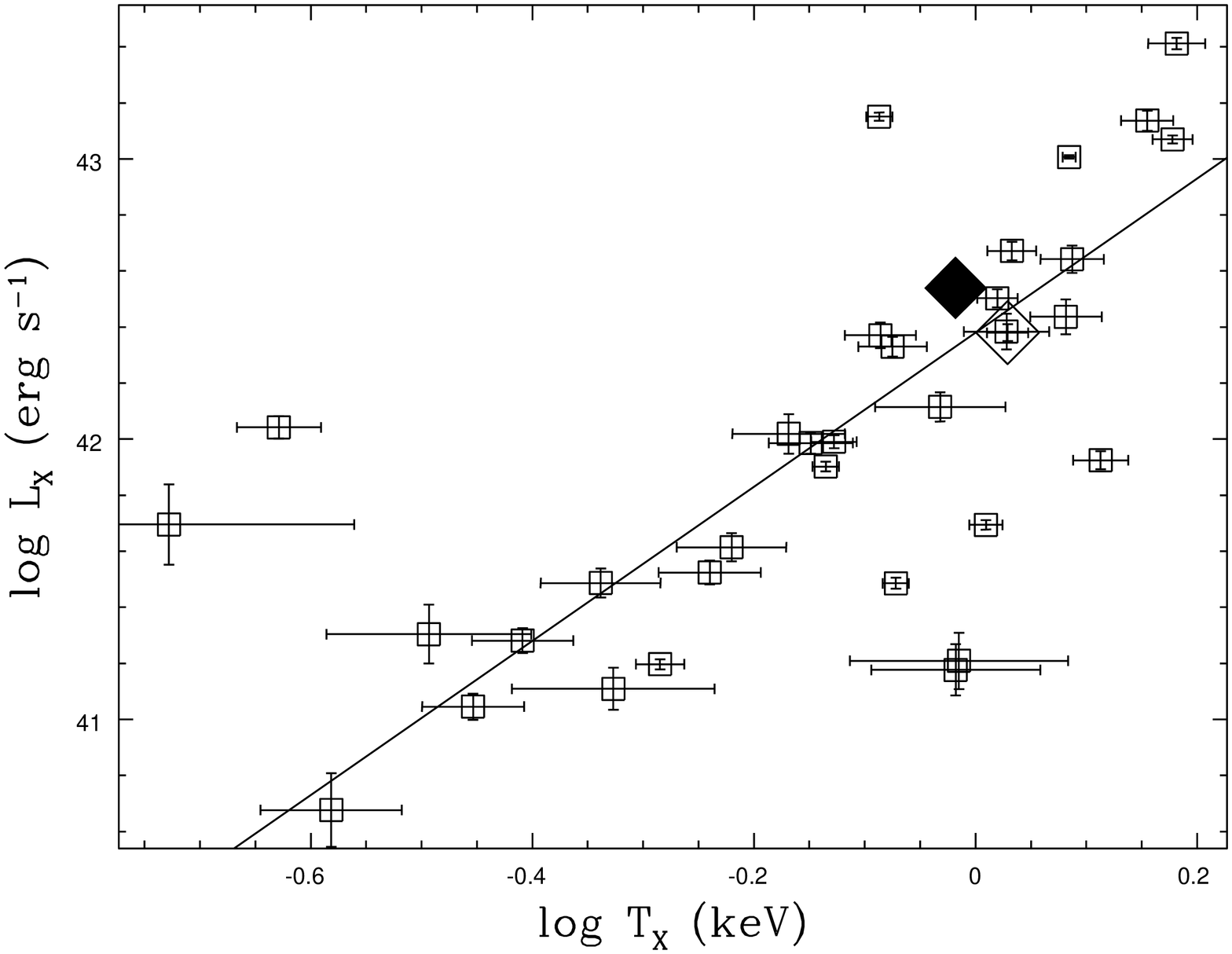}
  \includegraphics[angle=270,width=0.49\linewidth]{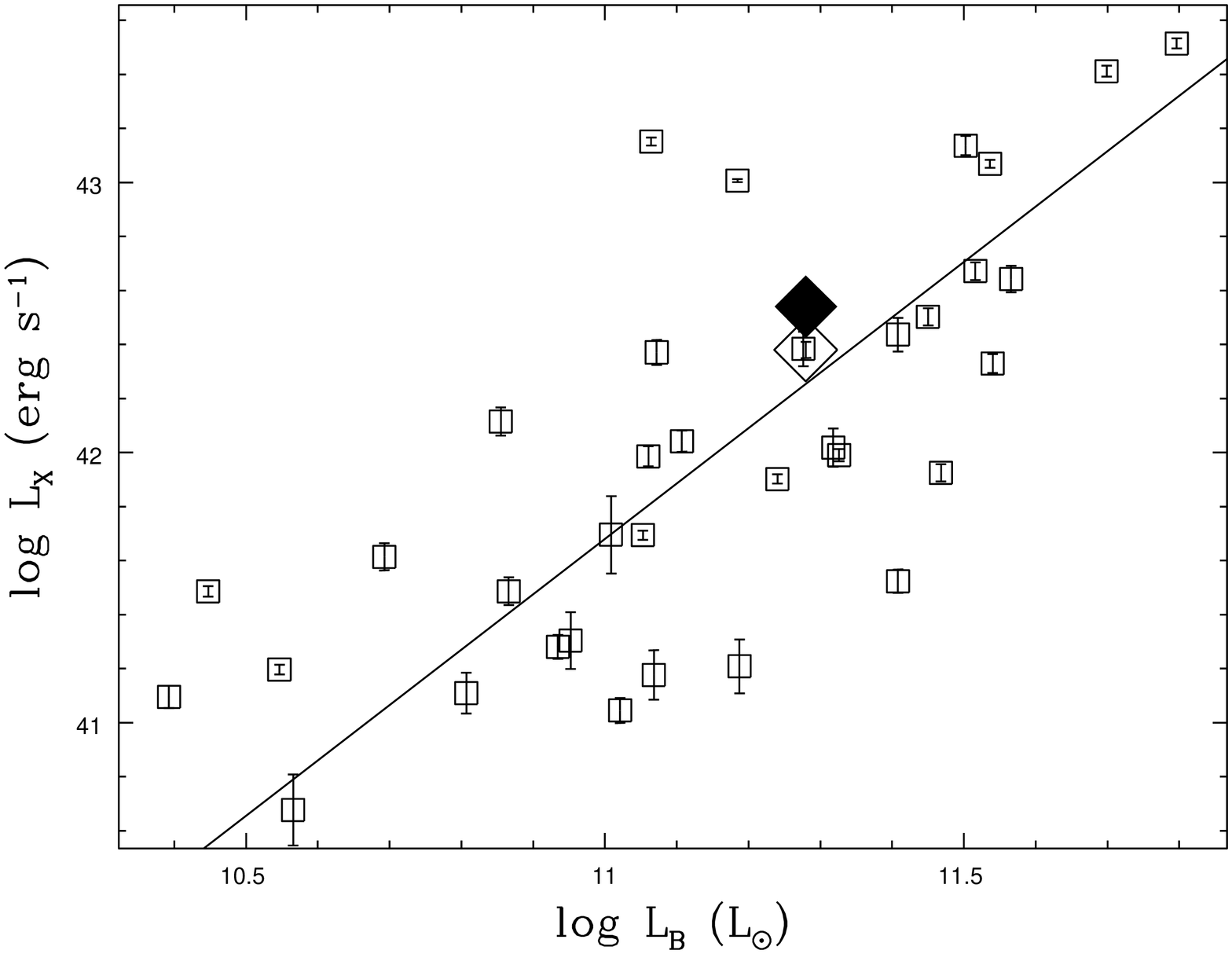}
  \includegraphics[angle=270,width=0.49\linewidth]{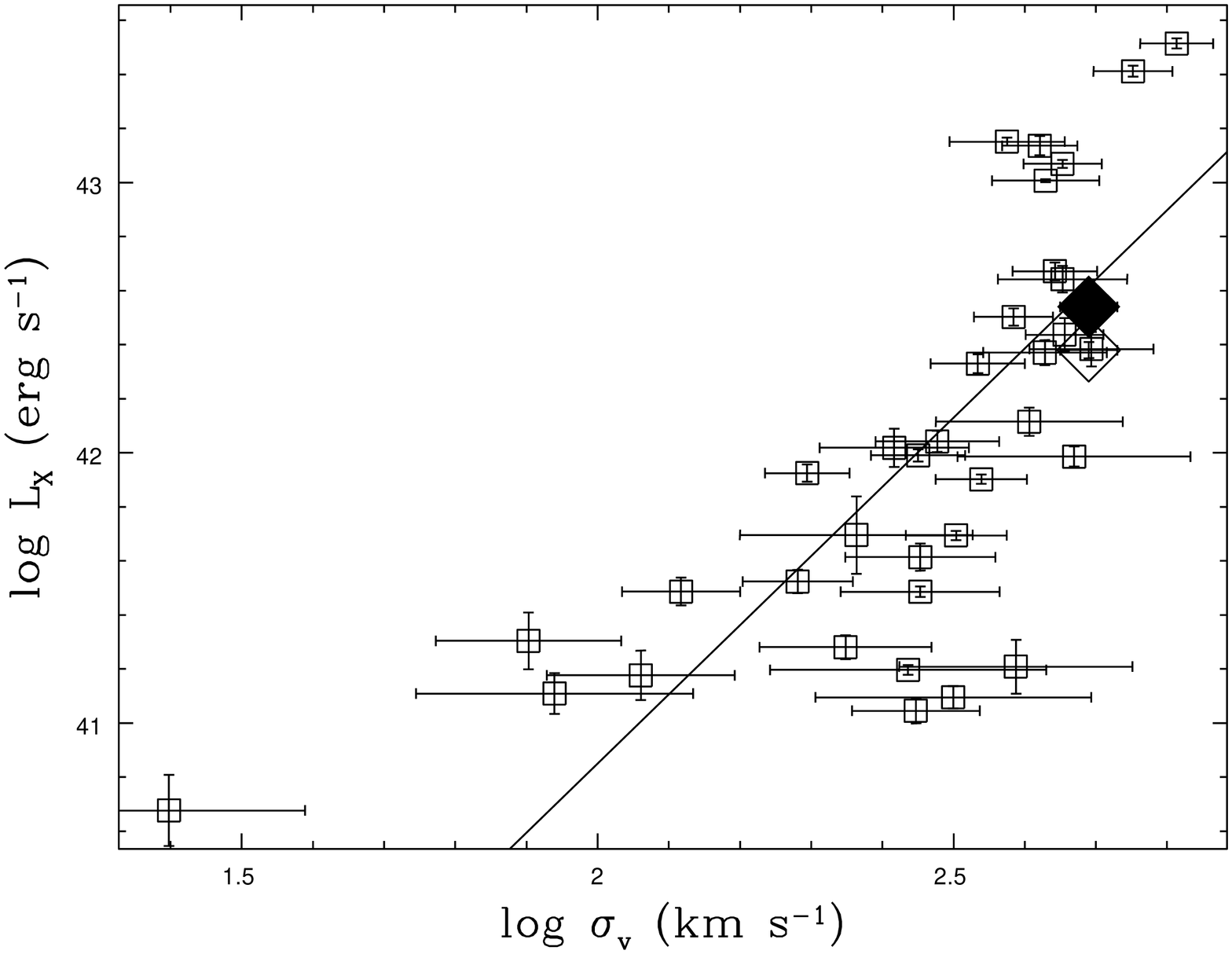}

  \caption{The scaling relations for galaxy groups as derived by
           \citet{osmond04a}.  The solid lines represent unweighted OLS bisector
           fits to the points shown, and the bold line in the \sigmav\ vs \TX\
           plot corresponds to \betaspec\ = 1.  The large diamonds represent
           NGC\,5171 from this work (solid), and \citeauthor{osmond04a} (open).}
  \label{fig_scaling}

\end{figure*}

Fig.~\ref{fig_scaling} shows the scaling relations derived by \citet{osmond04a},
with \ROSAT-derived values for NGC\,5171 shown as large open diamonds, and the
values derived in this work as filled diamonds.  The \sigmav-\TX\ plot shows the
high \betaspec\ of NGC\,5171 in comparison to most of the remainder of the \GEMS\
sample (\betaspec\ = 1 is indicated by the solid line).  The \LX-\TX\ and
\LX-\LB\ relations reveal that NGC\,5171 also has a rather high X-ray luminosity
in comparison to both its temperature and optical luminosity.  The final plot
shows the \LX-\sigmav\ relation in which NGC\,5171 appears consistent with the
line of best-fit for groups.  These results suggest that some effect is serving
to either increase \sigmav\ and \LX\ or decrease both \TX\ and \LB\ in the
NGC\,5171 group.

\subsection{Galaxy dynamics}

Fig.~\ref{fig_galhist} shows a velocity histogram for the member galaxies of
NGC\,5171, where the bins are of width 300 \kmps, and are positioned such that
one bin is centred on the average value of velocity for the group (\vel\ = 6924
\kmps).  This plot demonstrates weak evidence for a bimodal distribution in
galaxy velocities, which may be responsible for the high velocity dispersion seen
in NGC\,5171.  It should however be noted that the significance of this
bimodality is sensitive to the position of the histogram bins, and moving them
by $\sim$ 100 \kmps\ can remove this feature altogether.  However the
distribution remains non-Gaussian in appearance independent of the binning phase.
To measure the departure of the observed velocity histogram from a Gaussian
distribution (mean = \vgroup, sigma = \sigmav), we apply the Kolmogorov-Smirnov
test, which does not require the data to be binned, and is appropriate for a
small number of data points.  We find a probability of agreement between the two
distributions of $<$ 0.1\%, suggesting that NGC\,5171 exhibits genuine velocity
substructure.

The high values of \sigmav\ and \betaspec\ derived for NGC\,5171, together with
the non-Gaussian nature of its velocity histogram, suggests the presence of two
sub-populations of galaxies in the process of merging.  From Fig \ref{fig_galmap}
we can see that the galaxies which constitute these two sub-populations are not
clearly spatially separated, indicating that either they are merging in a
direction close to the line of sight, or are currently passing through one another.

\subsection{Spectral features}

It is likely that the spectral features seen in Figs~\ref{fig_specmap_adapt} and
\ref{fig_specmap_region} are a direct result of the interaction of these two
sub-populations of galaxies.  The two features of interest are the bright central
filament, and the Western hot region, both of which are located near the BGG
(NGC\,5171), and the two brightest galaxies from the second population
(NGC\,5179 and NGC\,5176, see Fig.~\ref{fig_galmap}).

The enhanced abundance and reduced temperature of the bright central filament
(Fig.~\ref{fig_specmap_region}) suggests that it originates from the
supernova-enriched interstellar medium (ISM) of a bright galaxy.  If we assume
that the two populations of galaxies, dominated by NGC\,5171 and NGC\,5179
respectively ($\triangle$\vel\ = 221 \kmps), are in the process of merging, then
it is probable that the bright central filament is the result of a tidal
interaction between NGC\,5171 and NGC\,5176.

\begin{figure}
  \centering

  \includegraphics[height=\linewidth,angle=270]{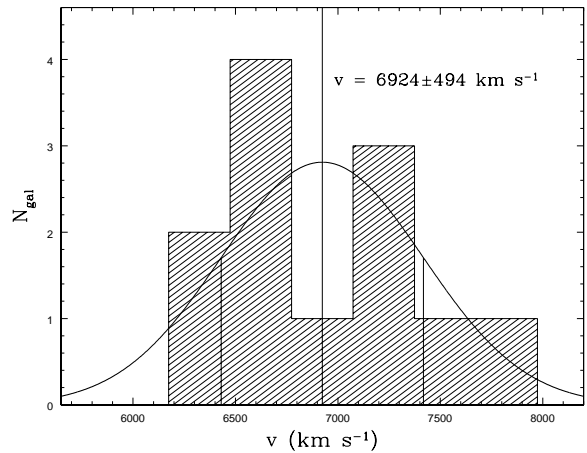}
  \caption{A velocity histogram of the member galaxies of NGC\,5171.  The curve
           represents the expected Gaussian distribution, using the calculated
           values of mean group velocity (\vel) and group velocity dispersion
           (\sigmav), marked with the intersecting vertical lines.}
  \label{fig_galhist}

\end{figure}

The lack of any increase in surface-brightness in the Western hot region suggests
no corresponding increase in density, and therefore an excess in entropy.  The
enhanced temperature and entropy of the this region could then result from
shock-heating of the IGM, by the interaction between the two merging subgroups.
The temperature of a region of shocked gas can be calculated from the velocity
difference between the two interacting systems using the standard shock jump
condition:

\begin{equation}
  \frac{\Tsh}{\TX} ~  = ~ \frac{[2\gamma \M ^{2}-(\gamma -1)][(\gamma -1)\M ^{2}+2]}{(\gamma +1)^{2} \M ^{2}}
  \label{eqn_Tsh}
\end{equation}

\noindent where \TX\ is the pre-shock temperature, $\gamma$ is the ratio of specific
heats (here 5/3), and \M\ is the Mach number of the shock.  We evaluate the spectral
properties of the hot region by extracting data from within the box shown in
Fig.~\ref{fig_specmap_adapt}, for all three instruments, and fitting in the usual
way.  This gives a temperature of 1.58$\pm$0.36 \kev, and a metallicity of
0.16$\pm$0.22 \Zsol.  Assuming the pre-shock temperature is equal
to the global value \TX\ = 0.96 \kev, implies a shock velocity of \vs\ =
641$\pm$132 \kmps.  From Fig.~\ref{fig_galhist} we see that this shock velocity
is consistent with the velocity separation between the two most populated bins
($\triangle$\vel\ = 600 \kmps).  It should also be noted that the value of
$\triangle$\vel\ = 600 \kmps\ is the velocity difference along the line of sight,
and is therefore a lower limit on the 3D velocity difference.

Hydrodynamical simulations of merging clusters of galaxies have shown that
shock-heating can occur at the interface of the two subclusters, and then
propagate out in a direction perpendicular to the axis of collision
\citep{takizawa99}.  In the case of a collision between two galaxies of unequal
mass, the shocked region forms a cone, angled towards the more massive galaxy.
Our observed hot region is therefore consistent with an interaction in the
direction of the bright central filament, where NGC\,5171 is the more massive
galaxy, and NGC\,5176 belongs to a less massive infalling group.  However there
is only weak evidence for a corresponding hot region to the South-East of this
filament (Fig.~\ref{fig_specmap_adapt}).

\subsection{South-East emission region}

Given the undetermined redshifts of the galaxies coincident with the South-East
emission region (Section \ref{sec_optical}), it is likely that the X-ray emission
seen here originates from a background group or cluster of galaxies.  We evaluate
the spectral properties of this region by extracting data from within a radius of
2 arcmin of the background object, and for the purposes of spectral fitting,
assuming a redshift 4 times greater than the redshift of  NGC\,5171, based on the
lower galaxy luminosities discussed in Section \ref{sec_optical}.  We find a
temperature of 1.14$\pm$0.13 \kev, and a metal abundance of 0.16$\pm$0.14 \Zsol.
At the distance of NGC\,5171, the X-ray luminosity corresponding to this model is
41.37$\pm$0.32 \ergps.  Adopting the \LX-\TX\ relation for groups from
\citet{osmond04a}, and using our fitted value of temperature, gives a predicted
X-ray luminosity of 42.54 \ergps.  Comparing these two \LX\ values suggests that
if the South-East emission originates from a galaxy group, its distance should be
a factor of $\approx$ 3.8 larger than that of NGC\,5171, in excellent agreement
with the factor of $\approx$ 4 derived from the galaxy luminosities (Section
\ref{sec_optical}).  We therefore conclude that the background object is
probably a group of galaxies at a distance of $\approx$ 430 \Mpc\ (z $\approx$
0.092).


\section{Conclusions}
\label{sec_conclusions}

Previous studies of NGC\,5171 using the \ROSAT\ \PSPC\ have found it to be both
X-ray luminous (\LX\ $>$ 10$^{42}$ \ergps), and of an unusual morphology.  X-ray
contours appear highly irregular and are not centred on a BGG
(Fig.~\ref{fig_overlay}).  However improved \XMM\ data reveal that the X-ray
emission is comprised of two distinct subregions corresponding to NGC\,5171 and
a background group, and that the main component is centred on a bright central
ridge extending from the BGG, to a nearby galaxy.  Spectral mapping reveals this
filament to be both cool and metallic, suggesting that it originates from the
interstellar medium of the central galaxies.  We therefore suggest that this
filament is the result of a tidal interaction between NGC\,5171 and NGC\,5176.

Our spectral analysis of NGC\,5171 indicates that the integrated X-ray emission
is reasonably well described by a single temperature APEC model, with \TX\ =
0.96 \kev, typical of groups \citep{osmond04a}.  However the calculated
values of \sigmav, and therefore \betaspec, are unusually high, and a histogram
of the galaxy velocities exhibits a significantly non-Gaussian distribution, as
well as a weak bimodality.  This suggests that NGC\,5171 may be comprised of two 
or more sub-populations of galaxies in the process of merging.

Adaptively binned spectral maps reveal a hot, metallic region to the West of the
bright central filament.  The temperature of this region is consistent with
shock-heating by galaxies moving with a velocity similar to the velocity
separation of the two sub-populations of galaxies.

We also compare our data with the \ROSAT\ derived values of \citet{osmond04a},
finding agreement within 1.1$\sigma$ in \TX, and 1.6$\sigma$ in \LX.  This
comparison helps to validate the use of \ROSAT\ data in deriving global group
properties, and hence scaling relations.

\section{Acknowledgements}

This research has made use of the $\NASA$-$IPAC$ Extragalactic Database, and
optical images from the STScI Digitised Sky Survey.  The XMM-Newton project is
supported by the Bundesministerium f\"ur Bildung und Forschung/Deutsches Zentrum
f\"ur Luft- und Raumfahrt (BMFT/DLR), the Max-Planck Society and the
Heidenhain-Stiftung, and also by PPARC, CEA, CNES, and ASI. The authors also
acknowledge the support of a studentship (JPFO) and a senior fellowship (TJP)
from the Particle Physics and Astronomy Research Council, and the
Verbundforschung grant 50 OR 0207 of the DLR (AF).



\label{lastpage}
\end{document}